\documentclass[conference]{IEEEtran}
\IEEEoverridecommandlockouts
\usepackage{cite}
\usepackage{amsmath,amssymb,amsfonts}
\usepackage{algorithmic}
\usepackage{graphicx}
\usepackage{textcomp}
\usepackage{balance}
\usepackage{hyperref}
\usepackage{xcolor}
\def\BibTeX{{\rm B\kern-.05em{\sc i\kern-.025em b}\kern-.08em
    T\kern-.1667em\lower.7ex\hbox{E}\kern-.125emX}}
\newtheorem{definition}{Definition}
\begin{document}

\title{Pre-training Graph Neural Network for Cross Domain Recommendation}
\author{\IEEEauthorblockN{Chen Wang\IEEEauthorrefmark{1},
Yueqing Liang\IEEEauthorrefmark{2},
Zhiwei Liu\IEEEauthorrefmark{1},
Tao Zhang\IEEEauthorrefmark{1}, and Philip S. Yu\IEEEauthorrefmark{1}}
\IEEEauthorblockA{\IEEEauthorrefmark{1}Department of Computer Science, University of Illinois at Chicago, USA}
\IEEEauthorblockA{\IEEEauthorrefmark{2}Business School, University of Sydney, Sydney, AU}
\IEEEauthorblockA{
\{cwang266, zliu213, tzhang90, psyu\}@uic.edu, \{ylia7737\}@uni.sydney.edu.au}}

\maketitle

\begin{abstract}
A recommender system predicts users' potential interests in items, where the core is to learn user/item embeddings. Nevertheless, it suffers from the data-sparsity issue, which the cross-domain recommendation can alleviate. However, most prior works either jointly learn the source domain and target domain models,  
or require side-features. However, jointly training and side features would affect the prediction on the target domain as the learned embedding is dominated by the source domain containing bias information. 
Inspired by the contemporary arts in pre-training from graph representation learning, we propose a pre-training and fine-tuning diagram for cross-domain recommendation. 
We devise a novel Pre-training Graph Neural Network for Cross-Domain Recommendation (PCRec), which adopts the contrastive self-supervised pre-training of a graph encoder. Then, we transfer the pre-trained graph encoder to initialize the node embeddings on the target domain, which benefits the fine-tuning of the single domain recommender system on the target domain. The experimental results demonstrate the superiority of PCRec. Detailed analyses verify the superiority of PCRec in transferring information while avoiding biases from source domains.
\end{abstract}


{\centering
\begin{figure*}[!hbt]
    \includegraphics[width=0.95\linewidth]{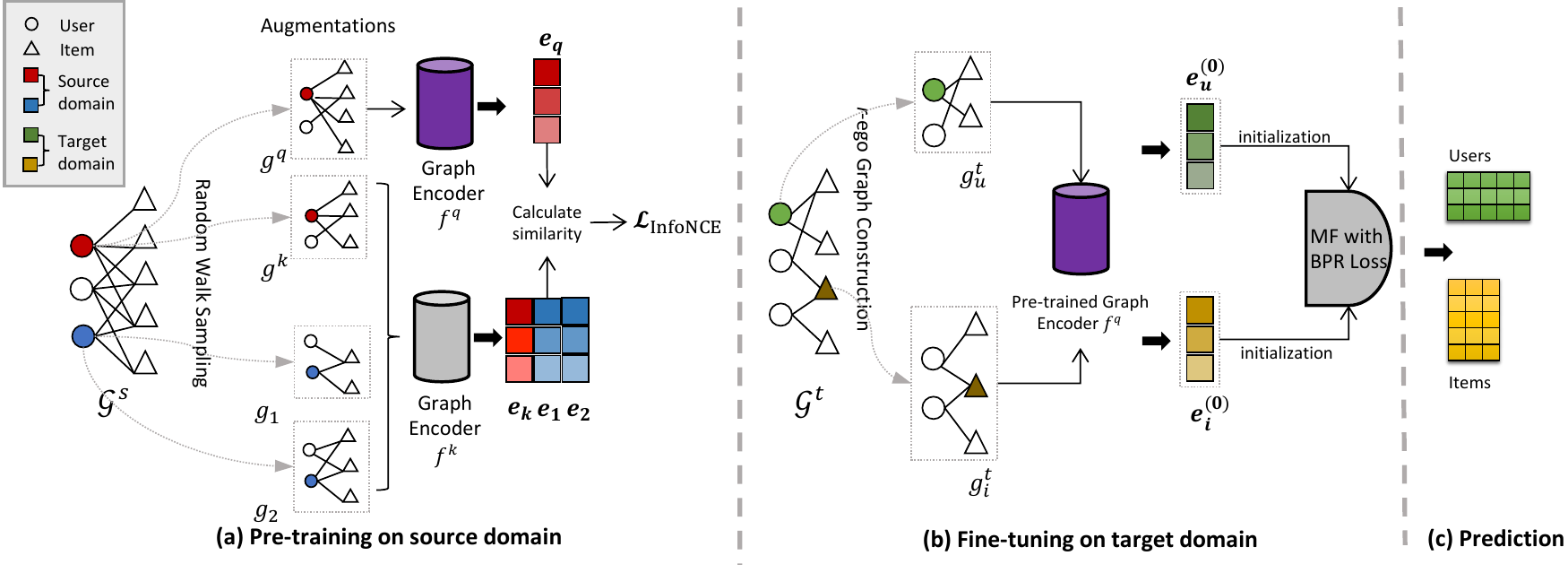}
    \caption{An illustration of PCRec model architecture in specific example. The model is mainly composed of three setps. In the pre-training stage, a model with the structural learning of nodes' embedding is used to learn the source domain's structure. The target domain is put into the pre-trained model to initialize nodes' embedding and fine-tune by a bipartite recommendation system in the fine-tuning stage. In the last, applied fine-tuned embedding to predict.
 }
\label{fig:flowchart}
\end{figure*}
}

\section{Introduction}
A recommender system predicts the potential interests of users to items, where the core is to learn user/item embeddings. Matrix factorization is an early method for collaborative filtering (CF), where it learns user/item embeddings and reconstructs their interactions with inner product~\cite{DBLP:journals/computer/KorenBV09, DBLP:conf/nips/SalakhutdinovM07}. The general idea of CF approach is based on observation of the preferences of other users that are similar to the historical preferences of the target user. 
Later on, deep recommender systems~\cite{DBLP:conf/kdd/WangWY15,liu2020basconv,DBLP:conf/www/HeLZNHC17,9377917} have shown that capturing deep features in a supervised
or unsupervised manner is more appealing than shallow models such as CF to capture similarity and implicit
relationship between items. However, due to the cold-start and data-sparsity issues~\cite{DBLP:journals/tkde/AdomaviciusT05, DBLP:conf/sigir/ScheinPUP02}, models learned from a single domain are unable to achieve satisfactory performance. Therefore, Cross-Domain Recommendation~(CDR)~\cite{DBLP:journals/csur/KhanIG17,DBLP:conf/ijcai/ManSJC17, DBLP:conf/sigir/FarseevSFC17, DBLP:conf/www/HuCXCGZ13, DBLP:conf/cikm/HuZY18} has been proposed. It transfers the information from source domains to a target domain, such that the recommendation performance on the target domain can be improved.

The CDR has been verified as an effective way to alleviate the aforementioned issues~\cite{DBLP:conf/ijcai/ManSJC17, DBLP:conf/sigir/FarseevSFC17, DBLP:conf/www/HuCXCGZ13, DBLP:conf/cikm/HuZY18}. The essential idea in CDR is to use the common users on both domains to transfer relevant information, which can be achieved from two perspectives: 1) modeling user information in source domain as auxiliary information in target domain~\cite{DBLP:conf/ijcai/ManSJC17, DBLP:conf/www/HuCXCGZ13}; or 2) jointly training shareable parameters on both domains~\cite{DBLP:conf/sigir/FarseevSFC17, DBLP:conf/dasfaa/WangPWYFH18}. Meanwhile, some recent works~\cite{DBLP:conf/cikm/HuZY18, DBLP:conf/wsdm/0008T20} try to combine both. However, existing works assume that the information from a source domain is relevant to the prediction objectives in the target domain, which is not necessarily true. If a source domain contains dominant bias against the target domain, the prediction on the target domain would be misled, which thus spoils the recommendation performance. 

To this end, we should devise a new CDR paradigm, which can not only transfer the information but also protect the prediction on the target domain from being dominated by the bias from source domains. Inspired by recent developments of pre-training frameworks in computer vision~\cite{DBLP:conf/cvpr/WuXYL18, DBLP:conf/cvpr/He0WXG20}, natural language processing~\cite{devlin2018bert}, graph representation learning~\cite{DBLP:conf/kdd/QiuCDZYDWT20}, and sequential recommendation~\cite{liu2021contrastive}, we intent to design a novel pre-training framework for the CDR problem. The advantages are twofold: Firstly, the pre-trained model on source domains transfer auxiliary information to the target domain. Secondly, the fine-tuning step on target domain ensures the prediction is dominated by the information in the target domain, thus overcoming the bias from source domains. 

Though existing works have discussed how to pre-train a model on source domains and fine-tune it on the target domain~\cite{Hu*2020Strategies}, it is still challenging to propose a suitable framework for the CDR problem. Firstly, previous works~\cite{DBLP:conf/kdd/QiuCDZYDWT20,DBLP:journals/corr/abs-1905-12265,devlin2018bert} pre-train a model on a large corpus with side information available, which is not the case for CDR. The only available data in CDR is the user-item interactions, which is very sparse. 
Additionally, it is unclear how to exert the pre-trained model while avoiding its bias impeding the prediction on the target domain.  

Therefore, we propose to \textbf{P}re-train a graph neural network for \textbf{C}ross-domain \textbf{Rec}ommendation (PCRec) by transferring the graph structural information from source domain graphs to the target domain. During the pre-training stage, we adopt the Self-Supervised Learning~(SSL)~\cite{li2021pretraining,wu2020self,you2020graph} scheme to train a graph encoder. SSL reduces the prediction bias from source domains as it pre-trains the model without prediction objectives. To be more specific, we sample and embed two sub-graphs of a node as positive pairs and employ contrastive learning to maximize the agreements between them. 

Then, we transfer the pre-trained graph encoder to the target domain. However, since the encoder is also dominated by the structural bias in the source domain, we thus exert the pre-trained encoder to initialize node embeddings for another single domain model. According to our empirical study, we recognize that adopting a simple Matrix Factorization (MF) model on the target domain during the fine-tuning stage~\cite{DBLP:conf/uai/RendleFGS09} significantly outperforms other complex models, such as LightGCN (LGCN)~\cite{DBLP:conf/sigir/0001DWLZ020}. We hypothesis that a simple MF retains a balance between source domain and target domain information, while other complex models over-emphasize the data on the target domain and thus diminish the source domain information. The contributions are as follows:

\begin{itemize}
    \item To the best of our knowledge, we are the first work investigating pre-training GNN to tackle the CDR problem.
    \item We design a novel framework to handle the CDR problem, which pre-trains a graph encoder on source domain to initialize the user/item embeddings in the target domain.
    \item Extensive experiments are conducted to complete CDR and analyze our proposed framework.
\end{itemize}

\section{Related Work}
\subsection{Cross-domain recommendation}
The cross-domain recommendation system mainly alleviates two major CF-based model bottlenecks: user/item data sparsity and the cold-start problem. The trained model may be over-fitting when user/item data is too sparsity to cover the universal data distribution. The cold start problem often exists when new users have no historical shopping records and the lack of features of new products. Both issues lead to a decline in the effectiveness of the recommendation system.

Conventional CDR has two approaches~\cite{DBLP:conf/ijcai/ZhuW00L021} to address the above problems: (1) content-based transfer, and (2) embedding-based transfer. \textit{Content-based transfer} mainly links different domains by identifying similar content information while no common users/items in this situation~\cite{DBLP:journals/ngc/WinotoT08, DBLP:journals/ijon/TanBQCC14}. In contrast, \textit{Embedding-based transfer} focuses on user/item relevance, such as multiple domains have common users or common items. This approach first trains the CF-based model (bayesian personalized ranking~\cite{DBLP:conf/uai/RendleFGS09}, neural collaborative filtering~\cite{DBLP:conf/www/HeLZNHC17}, etc.)to extract user/item embeddings and then transfer these embeddings through common or similar users/items across domains. In this work, we focus on embedding-based transfer. Unlike learning common users/items attributes, we adopt self-supervised contrastive learning and graph neural networks to learn users' structural embedding from the source domain and then transfer it to the targer domain.

\subsection{Contrastive learning}
GNNs models couple with contrastive learning to learn graph or node level representations without relying on supervisory data~\cite{DBLP:journals/corr/abs-2105-07342}. Then the trained model can transfer the learned representations to a priori unknown downstream tasks. In general, the contrastive learning method needs to create multiple views for each instance in the dataset through various data augmentations~\cite{DBLP:conf/icml/HassaniA20}. Two views are positive pairs, one is the original instance, and another is the augmented instance. We also need two negative views generated from the different instances. The ultimate goal of contrastive learning is to shorten the distance of the positive pair while pushing the negative sample away. Mutual information (MI) is often the measurement in contrastive learning.


\section{Proposed Model}
\subsection{Preliminary}
Following existing works~\cite{DBLP:conf/bigdataconf/LiuZZHY19,DBLP:conf/kdd/QiuCDZYDWT20}, we model the user-item interactions as a bipartite graph. We denote the graph as $\mathcal{G} = (\mathcal{U},\mathcal{I},\mathcal{E})$, where $\mathcal{U}$, $\mathcal{I}$ and $\mathcal{E}$ denote the set of users, items, and edges, respectively. Regarding the CDR problem, we denote a source domain graph as $\mathcal{G}^{(s)}$ and a target domain graph as $\mathcal{G}^{(t)}$. For each node,
we extract its context information from the $r$-ego network 
which is defined as:
\begin{definition}[An $r$-ego network.]
The $r$-hop neighbors for a node $u$ are defined as $\mathcal{S}_u = \{i:d(u,i)\leq r\}$ where $d(u,i)$ is the shortest path distance between $u$ and $i$ in the graph $\mathcal{G}$. The $r$-ego network of vertex $u$, denoted as $\mathcal{G}_u$, is a sub-graph composed by $\mathcal{S}_u$ and the corresponding edges between $\mathcal{S}_u$.
\end{definition}
Next, we present how to pre-train a graph encoder $f:\mathcal{G}\rightarrow \mathbb{R}^{d}$ upon the source domain graph $\mathcal{G}^{(s)}$ by adopting the self-supervised learning scheme~\cite{DBLP:conf/kdd/QiuCDZYDWT20,wu2020self}. 

\subsection{Pre-training on Source Domain}
We adopt the SSL scheme during the pre-training phase, which employs contrastive learning to optimize the graph encoder. Specifically, the SSL has three components: 1) the data augmentation, which constructs positive and negative sub-graph pairs of a node, 2) the graph encoder to embed the sub-graphs, and 3) the contrastive loss to optimize the encoder. 

\subsubsection{Data Augmentation}
Contrastive learning requires the construction of positive pairs and negative samples of a node. As the only available data in the source domain graph is the interactions, we construct positive pairs as two sub-graphs of one node.  The sub-graphs should share similar structure information to warrant two sub-graphs to be positive pairs. Therefore, we sample them from the $r$-ego network of a node.

We first conduct two random walks on node $u$'s $r$-ego network $\mathcal{G}_u$ (the superscript is ignored for simplicity), to generate two sub-graphs $g^{q}$ and $g^{k}$, which are regarded as a positive pair. After constructing positive sub-graph pairs for nodes, we treat those sub-graphs generated from different $r$-ego networks as negative samples. 
We demonstrate the sub-graph construction process in Figure~\ref{fig:flowchart}(a), where $g_{u}^{q}$ and $g_{u}^{k}$ are a positive pair since they are sampled from the same node. We use $g^{k^{-}}_{1}$ and $g^{k^{-}}_{2}$ to denote the negative samples, which are sub-graphs sampled from the $r$-ego network of another node. 

\subsubsection{GNN Encoder}
After retrieving those sub-graphs, we feed them into two graph encoders $f^{q}$ and $f^{k}$, which is illustrated in Figure~\ref{fig:flowchart}(a). We encode the sub-graph $g^q$ with graph encoder $f^q$, while encoding other sub-graphs with $f^k$. Correspondingly, we generate low-dimensional representative vectors $\mathbf{e}^{q}$ and $\mathbf{e}^{k}$ for the positive pair $g^{q}$ and $g^{k}$, respectively. In this work, we choose the Graph Isomorphism Network (GIN)~\cite{DBLP:conf/iclr/XuHLJ19} to be the graph encoder because GIN exhibits powerful ability in distinguishing a broad class of graphs~\cite{DBLP:conf/focs/BabaiK79}. In general, other GNN models can also be used as the encoder. We leave this study as future work.

\subsubsection{Contrastive Loss Function}
We adopt the contrastive loss InfoNCE~\cite{DBLP:journals/corr/abs-1807-03748} to self-supervisedly optimize the graph encoder, which maximizes the agreements between positive pairs. The InfoNCE loss is formulated as follows:
\begin{equation}
  \mathcal{L}_{\text{InfoNCE}}=-\log\frac{\text{exp}(\mathbf{e}_{q}^\intercal\mathbf{e}_{k}/\boldsymbol{\tau})}{\sum_{i=1}^{n} \text{exp}(\mathbf{e}_{q}^\intercal \mathbf{e}_{i}/ \boldsymbol{\tau})},
\label{eq:1}
\end{equation}
where $\boldsymbol{\tau}$ is the temperature hyper-parameter. 
Minimizing this objective is equivalent to maximizing the similarity between positive pairs, \textit{i.e.} $\mathbf{e}_{q}$ and $\mathbf{e}_{k}$, while minimizing the similarity between negative pairs, \textit{i.e.} $\mathbf{e}_{q}$ and $\mathbf{e}_{i}$ where $i\neq k $. In practice, we view those instances as a query embedding $\mathbf{e}_{q}$ and a set of key embeddings $\{\mathbf{e}_{i}\}|_{i=1}^{n}$. The contrastive loss looks up a single key (denoted by $\mathbf{e}_{k}$) that $\mathbf{e}_{q}$ matches in
the key set. 

In contrastive learning, maintaining the K-size look-up key set is essential. Intuitively, as the denominator in Eq.~(\ref{eq:1}) expresses, larger key set size leads to better sampling of the underlying data space. Due to the computational constraint, we adopt the MoCo~\cite{he2020momentum} training scheme, which maintains a dynamic set of keys with a queue and a moving-averaged encoder. MoCo is able to increase the key set size without additional backpropagation costs. Formally, if denoting the parameters of $f_k$ as $\theta_k$ and those of $f_q$ as $\theta_q$, MoCo updates $\theta_k$ as $\theta_k \leftarrow m\theta_k +(1-m)\theta_q$, where $m\in [0,1)$ is a momentum hyper-parameter.

\subsection{Fine-tuning on Target Domain}
\label{subsec: fine-tuning on target domain}
After obtaining the pre-trained GIN model from the source domain, we 
should transfer the model to the target domain. However, due to the pre-trained encoder is dominated by the structural bias of the source domain graph, directly fine-tuning the encoder in the target domain cannot avoid the interference of the bias from the source domain.  Instead,
we employ the pre-trained GIN model to initialize the node embeddings in the target domain. Then, we fine-tune a simple Matrix Factorization~(MF)~\cite{DBLP:conf/uai/RendleFGS09} model with initialized embeddings to infer the final embeddings of nodes. We illustrate this process as in Figure~\ref{fig:flowchart}(b). The fine-tuning on the target domain is optimized with the Bayesian Personalized Ranking (BPR) loss function, which is formulated as follows~\cite{DBLP:conf/uai/RendleFGS09}:
\begin{equation}
  \mathcal{L}_{\text{BPR}} = -\sum_{u=1}^{M} \sum_{i\in \mathcal{N}_u} \sum_{j\notin \mathcal{N}_u} \log \sigma(\hat{y}_{ui} - \hat{y}_{uj}) + \lambda\|\Theta\|^2
\end{equation}
where $\mathcal{N}_u$ denotes the set of items which are the neighbors of node $u$, $\hat{y}_{ui}$, and $\hat{y}_{uj}$ denote the rating of user $u$ on item $i$, and the rating of user $u$ on item $j$ individually.

Alternatively,
the MF-based recommeder system can be substituted by any other single domain recommendation models, e.g., LightGCN. However, the empirical results indicate that a simple MF-based model outperforms a complex signal domain recommender system.

\subsection{Recommendation}
\label{subsec: recommendation}

We will use the embedding optimized during the fine-tuning stage to make a further recommendation. We compute the score between a user and an item as: $\hat{r}_{ui} = \boldsymbol{e}_u^\intercal \boldsymbol{e}_i$, 
where $\mathbf{e}_{u}\in \mathbb{R}^{d}$ and $\mathbf{e}_{i}\in \mathbb{R}^{d}$ are the user and item embeddings, respectively. The score $\hat{r}_{ui}$ is used to rank those items for users. 
\begin{table}
  \centering
  \caption{Statistics of the datasets.}
  \label{tab:data}
  \begin{tabular}{l|cccc}
    \hline
    Data  & \verb|#| User & \verb|#| Item & Sparsity & Domain\\
    \hline
    Amazon-GGF & 31,230 & 40,648 & 99.96\verb|%| & Source  \\
    Amazon-PP & 14,180 & 4,970 & 99.80\verb|%| & Target\\
    \hline
  \end{tabular}
\end{table}

\section{Experiments}

\begin{table}
\caption{The overall comparison. The performance is measured in the target dataset Amazon-PP. }
\label{tab:performance}
  {\small\centering
  \begin{tabular}{l|cc|cc}
    \hline
    Model & {\small Recall@20} & {\small Recall@40}
          & MAP@20 & MAP@40\\
    \hline
    LGCN & \underline{0.1935} & \underline{0.3107} 
        & 0.0203 & 0.0248\\
    CMF & 0.0712 & 0.1329 & 0.0080 & 0.0109 \\
    CoNet & 0.1153 & 0.2314 & 0.0160 & 0.0161\\
    JSCN & 0.1329 & 0.2520
        & \underline{0.0263} & \underline{0.0307}\\
    PCRec & \textbf{0.2756} & \textbf{0.4445} & \textbf{0.0264} & \textbf{0.0329} \\
    \hline
     Impro. & 42.43\% & 43.06\% & 0.38\% & 7.17\%\\
    \hline
\end{tabular}}
\end{table}
We conduct experiments to respond the following research questions~(RQs):
\begin{itemize}
    \item \textbf{RQ1:} Can PCRec outperform existing single domain and cross-domain recommendation methods?
    \item \textbf{RQ2:} In the case of cross-domain, how far away neighbor information aggregation is helpful to represent the node embedding?
    \item \textbf{RQ3:} In CDR problem, how can we effectively transfer the information from source domain to target domain?
\end{itemize}

\subsection{Data}
\label{subsec: data}
We conduct experiments on two datasets from the Amazon Review Data (2018)\footnote{\href{https://nijianmo.github.io/amazon/index.html}{Available at https://nijianmo.github.io/amazon/index.html}.}: Grocery and Gourmet Food (Amazon-GGF) and Prime Pantry (Amazon-PP). Their statistics are shown in Table~\ref{tab:data}. For cross-domain models, we set Amazon-GGF as the source domain and Amazon-PP as the target domain, while for single domain model, we solely use the target domain data. Due to distinct sparsity on both datasets, we adopt the 5-core\footnote{\href{https://en.wikipedia.org/wiki/Degeneracy_(graph_theory)}{Each of the remaining users has at least k ratings}.} setting for Amazon-PP, and 10-core setting for Amazon-GGF. The number of common users between Amazon-PP and Amazon-GGF is 4,275.

\subsection{Experimental Settings}
\label{subsec: experimental settings}
\subsubsection{Baselines}

We compare our proposed PCRec with three cross-domain and one single-domain methods.  \textbf{CMF}~\cite{DBLP:conf/kdd/SinghG08}  is a matrix factorization-based cross-domain rating prediction model. \textbf{CoNet}~\cite{DBLP:conf/cikm/HuZY18}  and \textbf{JSCN} \cite{DBLP:conf/bigdataconf/LiuZZHY19} both are joint learning model with differernt ways to transfer one domian knowledge to the other. One single domain method is \textbf{LightGCN}, which devises a light graph convolution for training efficiency and generation ability.

\subsubsection{Evaluation Protocol}
We randomly split 80\% and 20\% of the interactions in the target domain as training and testing set, respectively. We randomly choose 10\% of the training data for validation during training. We use \textit{Recall}@${K}$ and \textit{MAP}@${K}$ to evaluate the top-${K}$ recommendation performance where ${K}=[20, 40]$.

\subsubsection{Hyper-parameter Settings}
In the pre-training, we apply Adam optimizer with a learning rate of 0.005, $\tau$ is 0.07, and MoCo (\textit{K}=512) momentum $m$ is 0.999. We change the learning rate to 0.001 in adam optimizer, and the early stopping strategy is the same as LightGCN. Furthermore, PCRec method inherits the optimal values of other shared hyper-parameters.

\subsection{Overall Comparison (RQ1)}
\label{subsec: overall comparison}
We compare PCRec with various baselines regarding the performance on the target domain Amazon-PP. The overall comparison is reported in Table \ref{tab:performance} and we have the following findings:
\begin{itemize}
    \item PCRec can significantly outperforms other methods on recall, e.g., achieving  42.43\% on Recall@20. against the second-best one. This is because PCRec can effectively transfer the information from source domains to target domain while protecting the recommendation on the target domain from being dominated by the source domain. The performance variant between MAP and Recall is because the number of positive samples in the data is small.
    \item Among those baselines, LGCN performs the best with respect to Recall@\{20,40\}, even better than CDR methods, which indicates the effectiveness of using GNN to learn node embeddings. Nevertheless, it is still much worse than PCRec since it is cannot transfer the information from the source domain.
    \item JSCN outperforms other baseline CDR methods, which shows the benefits of using spectral graph convolution to encode user-item interaction. However, it performs worse than PCRec, because it jointly learns the source and target domain embeddings, which leads to the interference of the noise in the source domain. 
\end{itemize}

\begin{table}
\caption{The influence of distance on neighbors aggregation. }

\label{tab:rq2}
  {\small\centering
  \begin{tabular}{l|cc|cc}
    \hline
    Variants & {\small Recall@20} & {\small Recall@40} 
          & {\small MAP@20} & {\small MAP@40}\\
    \hline
    PCRec-2hop & \underline{0.0329} & \underline{0.0663} & \underline{0.0044} & \underline{0.0064}\\
    PCRec-3hop & 0.0297 & 0.0613 & 0.0037 & 0.0039\\
    \hline
    PCRec-L3 & 0.1938 & \underline{0.3267} & 0.0200 & \underline{0.0259}\\
    PCRec-L1 & \underline{0.1996} & 0.3264 & \underline{0.0201} & 0.0258\\
    \hline
    PCRec & \textbf{0.2756} & \textbf{0.4445} & \textbf{0.0264} & \textbf{0.0329}\\
    \hline
\end{tabular}}
\end{table}

\subsection{Variants Analysis (RQ2)}
\label{subsec: variant analysis}

In this section, we aim to analyse the specific designs of our proposed PCRec framework regarding the neighbors aggregation.

At the pre-training stage, how to generate the subgraphs is critical. Regarding this point, two designs can be modified: 
\begin{itemize}
    \item The first one is the choice of $r$ concerning the $r$-hop neighbors. We change the $r$ from 2 to 3, and results are generated by applying the pre-trained GIN without fine-tuning to reflect the intrinsic effect. Intuitively, PCRec should perform better at 3-hop rather than 2-hop due to more neighbor information being included. However, as presented in Table~\ref{tab:rq2}, PCRec-2hop outperforms PCRec-3hop. We argue that this is because, with the distance increasing, the similarity between positive samples will decrease, thus hindering node representation learning. 
    \item The second design that can be changed is the data augmentation method. Usually, there are four ways to augment graph data, node dropping, edge dropping, random walk, and attribute masking~\cite{DBLP:conf/nips/YouCSCWS20}. The first three mechanisms do not require side information. Thus they can be adopted for the CDR problem. In PCRec, we generate the subgraphs through random walk, which can be supported by the graph structure assumption~\cite{DBLP:conf/kdd/LeskovecKF05}. The other two methods can be explored in the future.
\end{itemize}

For the fine-tuning stage, we study how the complexity of fine-tuning model will impact the performance. We change PCRec's MF to 1-layer LGCN~(PCRec-L1) and 3-layer LGCN~(PCRec-L3). There is no significant difference between PCRec-L1 and PCRec-L3, and they are both much worse than PCRec. We hypothesis that a complex model will over-emphasize the target domain, thus hindering knowledge transfer. Therefore, a simple model with elaborate initialization may be able to retain the balance.

\begin{table}
\caption{Exploration of various ways to transfer the information from source domain to target domain of CDR.}
\label{tab:rq3}
  {\small\centering
  \begin{tabular}{l|cc|cc}
    \hline
     & {\small Recall@20} & {\small Recall@40} 
          & {\small MAP@20} & {\small MAP@40}\\
    \hline
    LGCN & 0.1935 & 0.3107 & 0.0203 & 0.0248 \\
    Pre-Only & 0.0329 & 0.0663 & 0.0044 & 0.0064\\
    CU-PE & 0.1911 & 0.3198 & 0.0199 & 0.0258\\
    CU-PM & \underline{0.2277} & \underline{0.3880} & \underline{0.0227} & \underline{0.0293} \\
    \hline
    PCRec & \textbf{0.2756} & \textbf{0.4445} & \textbf{0.0264} & \textbf{0.0329}\\
    \hline
\end{tabular}}
\end{table}

\subsection{Transfer in CDR (RQ3)}
\label{subsec: transferincdr}
To explore the process of implementing knowledge transfer in CDR, we present how do we finally get to the PCRec framework by altering three key components step by step: 

\begin{itemize}
    \item We first examine the necessity of fine-tuning in CDR. We transfer the source knowledge by the pre-trained GIN model without fine-tuning, which is named Pre-Only. As shown in Table~\ref{tab:rq3}, the performance of Pre-Only is the worst, which implies that naively adopting the pre-trained model would bring severe bias to the target domain. Therefore, we add a fine-tuning model to PCRec. 
    \item Secondly, we consider two plans to transfer the knowledge to common users: 1) straightly using the pre-trained embeddings as the initialization of fine-tuning (named CU-PE), 2) applying the pre-trained model to generate the initialization of fine-tuning (named CU-PM). As presented in Table~\ref{tab:rq3}, CU-PM not only performs better than CU-PE but also surpasses the LGCN, while CU-PE falls behind it. In the CDR problem, the result indicates that transferring the inherited structure information from the model has better performance than copying the common users' embedding from the source domain to the target domain directly. 
    \item Finally, we explore the role of nodes other than common users, which leads to our final PCRec. In CDR problem, other nodes should be carefully processed. But as they can also receive the information transferred by the pre-trained encoder, we suggest that they should be included. The experiment result demonstrates that the target performance would not be dominated by the source domain in this way of transferring. 
\end{itemize}

\section{CONCLUSION}
\label{sec: conclusion}
In this work, we investigate the possibility of pre-training a GNN to transfer structural representations in the source domain to the target domain to address the cold-start problem in the CDR task. We propose a novel framework, PCRec, which pre-trains a graph encoder to learn node embeddings from the source domain and apply it to the target domain to initiate embedding. A simple MF method during the fine-tuning stage can significant outperform other complicated methods on all metrics. 

\section*{Acknowledgments}
This work is supported in part by NSF under grants III-1763325, III-1909323,  III-2106758, and SaTC-1930941.

\bibliographystyle{IEEEtran}
\balance
\bibliography{PCRec}
\end{document}